\documentclass[12pt]{article}
\usepackage{oldgerm}
\usepackage{yfonts}
\usepackage{euler}
\usepackage{graphics}
\usepackage{bm}
\usepackage{graphicx}
\usepackage{amsmath}
\usepackage{amssymb}
\usepackage{amstext}
\usepackage{amscd}
\usepackage{amsfonts}
\usepackage{appendix}
\usepackage{wasysym}
\usepackage{epstopdf}
\usepackage{color}
\DeclareMathAlphabet{\EuFrak}{U}{euf}{m}{n}
\DeclareMathAlphabet{\EuScript}{U}{eus}{m}{n}

\newcommand{\nd}{\noindent}

\newcommand{\be}{\begin{equation}}
\newcommand{\ee}{\end{equation}}
\newcommand{\ben}{\begin{eqnarray}}
\newcommand{\een}{\end{eqnarray}}

\title{{\bf In defense of Tsallis' original probability distribution }}
\author{{A. Plastino$^1$ and M. C. Rocca$^1$} \\
\small{$^1$ La Plata National University and
Argentina's National Research Council}\\
\small{(IFLP-CCT-CONICET)-C. C. 727, 1900 La Plata - Argentina}}

\date{\today}

\begin{document}

\maketitle

\begin{abstract}

\nd \small{Tsallis' pioneer q-probability distribution \newline 
$P_i=\frac {[1+\beta(1-q)U_i]^{\frac {1} {q-1}}} {Z}$\newline $Z=\sum\limits_{i=1}^n [1+\beta(1-q)U_i]^{\frac {1}
{q-1}}$ \newline [J. of Stat. Phys., {\bf 52}
(1988) 479] has been recently attacked in ArXiv:1705.01752, in a Reply to our 
arXiv: 1704.07493 publication. 
 We show here that such an attack is groundless.}

\nd Keywords: MaxEnt, Tsallis' functional variation, q-statistics.

\end{abstract}

\renewcommand{\theequation}{\arabic{section}.\arabic{equation}}

\setcounter{equation}{0}

\section{The Tsallis'probability distribution}

We reply here to reference (arXiv:1705.01752) of Oikonomou and Bagci (OB by short)  \cite{1}.
Under the pretense of replying to our reference \cite{o}, they question our variational procedure in such paper, but in so doing they are really attacking   Tsallis' original probability distribution (PD) \cite{tsallis88}. Let us see first how  we proceeded in \cite{o}.  The pertinent variational equation is (OB's Eq. (1))

\begin{equation}
\label{eq3.4} \left(\frac {q} {1-q}\right)P_i^{q-1} +\lambda_1 U_i
+\lambda_2=0,
\end{equation}
and OB  call this  PD $P$ by the name PR1. Of course,  Eq. (\ref{eq3.4}) is the  Tsallis'  Euler-Lagrange one 
of \cite{tsallis88}. 

\vskip 2mm \nd To make things transparent,  we revisit now the procedure given in  \cite{o}. One first gives the Lagrange multipliers $\lambda_1$ and $\lambda_2$ a prescribed {\it form} in terms of a (thus far unknown) quantity $Z$: 
\begin{equation}
\label{eq3.6} \lambda_1=\beta q  Z^{1-q},
\end{equation}
\begin{equation}
\label{eq3.7}
\lambda_2=\frac {q} {q-1}Z^{1-q},
\end{equation}
and then has
\begin{equation}
\label{eq3.8}
P_i=\frac {[1+\beta(1-q)U_i]^{\frac {1} {q-1}}} {Z},
\end{equation}
so that normalization demands for $Z$ 
\begin{equation}
\label{eq3.9}
Z=\sum\limits_{i=1}^n [1+\beta(1-q)U_i]^{\frac {1}
{q-1}}.
\end{equation}
The ensuing PD is (curiously)  called PR2 by OB \cite{1}. 

\vskip 2mm \nd {\it It is obvious that PR1 and PR2 are one and the same PD! However,  OB claim that they are 
different. OB try to validate such strange statement with a graph of three curves.}

\vskip 2mm \nd They introduce still a third PDF that they call OB, and hypothetically follows from their own variational equation (called by them Eq. (5)). 
In such Eq. (5) they inadvisedly FIX the energy-Lagrange multiplier as $\beta$, with disastrous consequences, as we will presently show. From their variational equation one obtains for the PD:

\be (P_i)_{OB}= [(\frac{1-q} {q})(\beta U_i + \gamma)]^{1/(q-1)},  \ee
so that OB's normalization entails

\be \sum_i \, (P_i)_{OB} = 1 = \sum_i \,  [(\frac{1-q} {q})(\beta U_i + \gamma)]^{1/(q-1)},   \ee
and one immediately appreciates the sad fact that $\gamma$ cannot be obtained in closed form. This makes normalization a difficult task, particularly in the continuum limit. OB preposterously claim that their 
 $(P_i)_{OB}$ is identical to PR1, which is patently absurd.
 
\vskip 2mm

In order to get out of this conundrum OB state(see  below their graph) that things are remedied by setting their Lagrange multipliers $\beta,\,\,\gamma$  equivalent to ours  $\lambda_1,\,\,\lambda_2$ via

\be \beta =  -\lambda_1; \,\,\, \,\,\gamma = - \lambda_2.\ee But then, PR1 becomes
 identical to  $ (P_i)_{OB}$! These two PDFs cannot yield different results, as OB enthusiastically  and with absolute confidence claim.

\setcounter{equation}{0}

\section{The Renyi probability distribution}

It is asserted in \cite{pre} that Renyi's probability distribution (PD) is
\begin{equation}
\label{eq2.1}
P_i=\frac {[1+\beta(1-q)(U_i-<U>)]^{\frac {1} {q-1}}} {Z},
\end{equation}
with
\begin{equation}
\label{eq2.2}
Z=\sum\limits_{i=1}^n [1+\beta(1-q)(U_i-<U>)]^{\frac {1}
{q-1}}.
\end{equation}
It is erroneously stated in \cite{1} that, in the limit $q\rightarrow 1$, the above partition function $Z$ 
 becomes 
\begin{equation}
\label{eq2.3}
Z=e^{\beta<U>}\sum\limits_{i=1}^n e^{-\beta U_i}.
\end{equation}
This happens because the authors of \cite{1}  did not bother to take the limit of the complete PD.
 Indeed, from
\begin{equation}
\label{eq2.4}
P_i=\frac {[1+\beta(1-q)(U_i-<U>)]^{\frac {1} {q-1}}} 
{\sum\limits_{i=1}^n [1+\beta(1-q)(U_i-<U>)]^{\frac {1}
{q-1}}},
\end{equation}
one deduces that for $q\rightarrow 1$ one has
\begin{equation}
\label{eq2.5}
P_i=\frac {e^{-\beta(U_i-<U>)}} 
{\sum\limits_{i=1}^n e^{-\beta(U_i-<U>)}},
\end{equation}
or, equivalently, 
\begin{equation}
\label{eq2.6}
P_i=\frac {e^{\beta<U>}e^{-\beta U_i}} 
{e^{\beta<U>}\sum\limits_{i=1}^n e^{-\beta U_i}}.
\end{equation}
Thus, 
\begin{equation}
\label{eq2.7}
P_i=\frac {e^{-\beta U_i}} 
{\sum\limits_{i=1}^n e^{-\beta U_i}}, 
\end{equation}
and then
\begin{equation}
\label{eq2.8}
Z=\sum\limits_{i=1}^n e^{-\beta U_i}.
\end{equation}
We see that  (\ref{eq2.3}) from \cite{1} is not correct.

\section{Conclusion}  

In view of these considerations, one concludes that reference \cite{1}  has no logical support.

\end{document}